\documentclass[runningheads]{llncs}

\usepackage[T1]{fontenc}
\usepackage{graphicx}
\usepackage{booktabs}
\usepackage{tabularx}
\usepackage{array}
\usepackage{microtype}
\usepackage{amssymb}
\usepackage{hyperref}
\usepackage{url}
\usepackage{tikz}
\usetikzlibrary{
  arrows.meta,
  shapes.geometric,
  shapes.symbols,
  positioning,
  fit,
  backgrounds,
  calc
}

\definecolor{clientborder}{RGB}{100,100,100}
\definecolor{clientbg}{RGB}{245,245,245}
\definecolor{gatewayborder}{RGB}{70,130,180}
\definecolor{gatewaybg}{RGB}{219,234,247}
\definecolor{gatewaycomp}{RGB}{219,234,247}
\definecolor{farmerborder}{RGB}{34,139,34}
\definecolor{farmerbg}{RGB}{212,237,212}
\definecolor{farmercomp}{RGB}{212,237,212}
\definecolor{instborder}{RGB}{178,34,34}
\definecolor{instbg}{RGB}{245,210,210}
\definecolor{instcomp}{RGB}{245,210,210}
\definecolor{sharedborder}{RGB}{148,103,189}
\definecolor{sharedbg}{RGB}{235,225,245}
\definecolor{sharedcomp}{RGB}{235,225,245}
\definecolor{databorder}{RGB}{70,130,180}
\definecolor{databg}{RGB}{219,234,247}
\definecolor{datacomp}{RGB}{219,234,247}
\definecolor{extborder}{RGB}{210,140,30}
\definecolor{extbg}{RGB}{252,240,210}
\definecolor{extcomp}{RGB}{252,240,210}
\definecolor{arrowgray}{RGB}{100,100,100}
\definecolor{lblgray}{RGB}{120,120,120}
\definecolor{rlsborder}{RGB}{70,130,180}
\definecolor{rlsbg}{RGB}{235,245,255}

\newcolumntype{L}[1]{>{\raggedright\arraybackslash}p{#1}}

\begin{document}
\title{Governance-Aware Software Architecture for
Multi-Stakeholder Platforms}
\titlerunning{Governance-Aware Software Architecture for MSPs}

\author{
  Michael Nwankwo\inst{1} \and 
  Eric Umuhoza\inst{1}
}

\institute{
  Carnegie Mellon University, CMU-Africa, Kigali, Rwanda \\
  \email{cmnwankw@alumni.cmu.edu, eumuhoza@andrew.cmu.edu} 
}

\maketitle

\begin{abstract}
Multi-stakeholder platforms (MSPs) coordinate diverse
stakeholder groups, often with competing or conflicting
requirements. As these platforms increasingly take digital
form, engineers building them make architectural decisions
about data visibility, service decomposition, and algorithm
design that directly determine which stakeholder requirements
are prioritized when conflicts arise. Software architecture
literature provides patterns for data isolation and access
control among tenants but does not address how architectural
decisions resolve conflicts among stakeholders with
structurally divergent interests. MSP governance literature
identifies the principles at stake but treats technology as
neutral infrastructure. Neither addresses the translation
between governance principles and architectural decision
spaces. This paper proposes a governance-architecture
correspondence framework that surfaces implicit governance
decisions, making them explicit and debatable before
deployment. The framework maps five MSP governance principles
to the architectural decision spaces where they must be
addressed, identifying for each the governance-aware design
choice and the technically convenient default it overrides.
We illustrate the framework in a constructed knowledge
platform for pig farming in Rwanda, where five stakeholder
types present structurally conflicting requirements. As work
in progress, the framework is proposed but not yet empirically
validated; a planned pre/post judgment study with platform
users across all stakeholder types will test falsifiable
predictions about governance outcomes.
\end{abstract}

\section{Introduction}
\label{sec:intro}

Engineers building knowledge platforms for multi-stakeholder
coordination face a decision problem about whose requirements
are prioritized when stakeholder interests conflict. When a
government agency requires disease surveillance data and the
farmers being surveilled require protection from its
consequences, the engineer must decide where to enforce data
isolation. When institutional actors require authority
recognition and the platform is meant to value peer and local
knowledge equally, the engineer must decide how reputation
scoring assigns weight across signal types. When farmer-facing
features require optimization for low-bandwidth rural
connectivity and institutional dashboards require rich
analytics, the engineer must decide where to draw service
boundaries. These decisions are made in every multi-stakeholder
platform that becomes software.

Classical software architecture patterns provide mechanisms for
the technical aspects of these decisions. Multi-tenancy
patterns address data isolation between customers who share
infrastructure~\cite{bezemer2010}. Role-based access control
determines which users can perform which
operations~\cite{ferraiolo1992}. Service decomposition patterns
address modularity and independent
deployment~\cite{lewis2014,newman2021}. These patterns are
agnostic to stakeholder configuration: they can be applied to
any pattern of users and roles. What is missing is a reasoning
framework for cases where stakeholder groups have structurally
divergent interests, where one group's transparency requirement
is another group's surveillance threat. The mechanisms exist;
architectural engagement with stakeholder conflict as a
distinct concern does not.

This decision problem is not new in form.
Winner~\cite{winner1980} established that technical artifacts
embody political choices: the structures and systems engineers
build embody specific forms of power and authority whether
engineers intend them to or not. Friedman et
al.~\cite{friedman2006} extended this insight to information
systems, showing that design decisions from database schemas
to interface flows encode values affecting stakeholder
autonomy and power. The implication for multi-stakeholder
platforms is that engineers make governance decisions whether
they recognize them as such or not: the technical choice and
the governance consequence are the same decision.

MSP governance literature identifies what is at stake when
these embedded decisions go unexamined. Van Ewijk et
al.~\cite{vanewijk2024} distill five design principles for
farmer-centered MSPs: safe spaces for participation without
surveillance, farmer-driven agendas, valuing innovations from
below, cross-level coordination without dominance, and
transparency for trust. Hermans et al.~\cite{hermans2017}
demonstrate empirically that structural properties of platform
design determine whether inclusive innovation or reproduced
hierarchy results. This literature provides robust governance
principles but treats technology as neutral infrastructure.
The gap this paper addresses is the translation between MSP
governance principles and software architectural decision
spaces: existing literature provides one or the other, but
not the connection between them.

This paper proposes a governance-architecture correspondence
framework that makes this translation explicit, illustrated in
a constructed knowledge platform for pig farmers in Rwanda.
We contribute:
\begin{enumerate}
 
  \item A mapping framework that connects five MSP governance
    principles to the architectural decision spaces where they
    must be addressed, identifying for each the
    governance-aware design choice and the technically
    convenient default it overrides.
  \item An illustrative instantiation in a constructed platform
    for Rwandan pig farming, documenting how each mapping was
    operationalised.
  \item A research agenda with falsifiable predictions
    structuring planned empirical validation through a pre/post
    judgment study with platform users across all stakeholder
    types.
\end{enumerate}
As work in progress, this paper does not claim that
governance-aware architectural choices produce better
governance outcomes; that requires the empirical evaluation
outlined as contribution (3).
\section{Background and Gap}
\label{sec:background}

MSP governance literature identifies principles critical to
inclusive multi-stakeholder coordination: creating safe spaces
for participation without surveillance, ensuring farmer-driven
agendas, valuing innovations from below, enabling cross-level
coordination, and building transparency for
trust~\cite{vanewijk2024}. Social network analysis of MSPs in
Rwanda, Burundi, and DRC demonstrates that structural
properties, not formal governance rules, determine whether
platforms achieve inclusive innovation or reproduce existing
hierarchies~\cite{hermans2017}. Innovation platforms that fail
to embed these arrangements in their structure reproduce
existing patterns rather than transforming
them~\cite{schut2016}. This literature treats technology as
neutral infrastructure; the translation from governance
principles to architectural decisions remains unaddressed.

Software architecture literature addresses the technical
challenge of serving multiple user groups from shared
infrastructure. Microservices patterns enable independent
deployment and team autonomy~\cite{lewis2014,newman2021}.
Multi-tenancy patterns isolate data between
tenants~\cite{bezemer2010}. Role-based access control governs
permitted operations~\cite{ferraiolo1992}. These patterns
assume tenants are customer organizations with similar needs
differing primarily in their data. Governance means access
control. The distribution of voice, power, and benefit that
MSP literature addresses does not appear.

The gap is consequential. Engineers building multi-stakeholder
platforms make governance decisions whether they recognize them
as such or not. Choosing a shared database over tenant
isolation decides who can see whose data. Designing a
reputation algorithm decides whose knowledge counts. Drawing
service boundaries decides whose requirements are independently
optimized when they conflict. Classical patterns applied
without governance reasoning reproduce the power asymmetries
that MSP literature warns against, not through malice but
through unreflective application of defaults that assume
stakeholder homogeneity. The framework in
Section~\ref{sec:framework} makes these decisions explicit
and debatable.

\section{Governance-Architecture Correspondence Framework}
\label{sec:framework}

\subsection{Framework Overview}

The framework rests on three premises. First, governance
principles imply architectural requirements: each MSP principle
can be decomposed into requirements that platform architecture
must satisfy. Safe spaces implies data isolation between farmer
contributions and institutional dashboards. Valuing innovations
from below implies reputation mechanisms that do not require
expert gatekeeping for peer knowledge to accumulate standing.

Second, architectural decisions have governance consequences.
A row-level security policy is simultaneously a data protection
mechanism and a power distribution mechanism. A service
boundary is simultaneously a modularity decision and a
stakeholder boundary decision. These are not two separate
decisions; they are one decision with two kinds of consequences.

Third, making implicit assumptions explicit enables
deliberation. A developer building a government dashboard who
queries farmer posts directly is not making a neutral technical
choice: they are enabling individual-level surveillance of the
actors the platform is meant to protect. The framework surfaces
these decisions before implementation rather than discovering
their consequences after deployment.

A reasonable objection to this framing is that row-level
security, microservices decomposition, and multi-signal
reputation scoring are conventional mechanisms rather than
novel architectural patterns. We acknowledge this: the
framework does not introduce new mechanisms. Its contribution
is the explicit linking of stakeholder-conflict reasoning to
where in architecture each conflict gets resolved. The same
database mechanism encodes different governance regimes
depending on how it is configured: a shared database with
role-based application-level access encodes one regime;
row-level security restricting institutional roles to
aggregates encodes another. Both are technically valid
configurations of the same conventional mechanisms; they
encode different choices about whose participation the
platform protects. The framework's value lies in making this
choice a deliberate, debatable architectural decision rather
than an unreflective default.

\subsection{The Governance-Architecture Mapping}

The five principles mapped in Table~\ref{tab:mapping} are drawn from van Ewijk et al.~\cite{vanewijk2024} and Hermans et al.~\cite{hermans2017} as a coherent set developed within the same agricultural MSP tradition this paper engages. The MSP literature contains other formulations of governance concerns; Schut et al.~\cite{schut2016}, for example, foreground institutional embedding as the central determinant of platform outcomes. We selected this principle set because its authors specified the requirements each principle implies in terms concrete enough to be addressed in architecture, which is the input the framework requires. Whether the same method, applied to other principle sets in other domains, produces useful mappings is an empirical question we do not resolve in this paper.

Table~\ref{tab:mapping} presents the mapping for these five
principles. For each principle, the table identifies the
governance requirement it implies, the architectural decision
space where that requirement must be addressed, and the
contrast between the governance-aware design choice and the
technically convenient default it overrides.

\begin{table}[t]
\caption{Governance-Architecture Mapping Framework}
\label{tab:mapping}
\scriptsize
\setlength{\tabcolsep}{4pt}
\begin{tabularx}{\textwidth}{@{}L{2.4cm}L{2.2cm}L{1.8cm}L{4.2cm}@{}}
\toprule
\textbf{Governance Principle} &
\textbf{Governance Requirement} &
\textbf{Decision Space} &
\textbf{Governance-Aware Design / Unaware Default} \\
\midrule

Safe spaces~\cite{vanewijk2024} &
Farmers participate without surveillance by powerful actors &
Data visibility &
\textit{Aware:} RLS isolating farmer content; aggregates only
for government \newline
\textit{Default:} Shared DB; institutional users access
individual posts \\

\addlinespace

Farmer-driven agenda~\cite{vanewijk2024} &
Farmers set priorities, not just consume content &
Content service design &
\textit{Aware:} Open Q\&A; community voting surfaces
priorities \newline
\textit{Default:} Curated expert library; farmers consume
but do not contribute \\

\addlinespace

Innovations from below~\cite{vanewijk2024} &
Peer and local knowledge valued alongside expert knowledge &
Reputation algorithm &
\textit{Aware:} Multi-signal scoring: peer validation
and expert endorsement \newline
\textit{Default:} Expert-only validation; content invisible
until approved \\

\addlinespace

Cross-level coordination~\cite{hermans2017} &
Local and institutional actors engage without one group
dominating &
Service decomposition &
\textit{Aware:} Stakeholder-aligned services; independent
optimization \newline
\textit{Default:} Monolith; lowest-common-denominator UX
for institutional needs \\

\addlinespace

Transparency for trust~\cite{friedman2006} &
Users understand how algorithmic systems affect them &
Algorithm visibility &
\textit{Aware:} Explicit reputation formula; explanatory
tooltips \newline
\textit{Default:} Black-box ranking optimized for engagement \\

\bottomrule
\end{tabularx}
\end{table}

\section{Instantiation: Rwanda Pig Farming Platform}
\label{sec:instantiation}

\subsection{Context and Stakeholder Conflicts}

The framework is illustrated in a knowledge platform for pig
farmers in Rwanda. Pig farming in Rwanda faces persistent challenges from disease outbreaks, limited veterinary access, and limited extension services~\cite{hirwa2022}. Rwanda's agricultural
extension system coordinates government agencies including
the Rwanda Agriculture and Animal Resources Development Board
(RAB), veterinary services, NGOs, and market actors,
exemplifying the multi-stakeholder coordination structure the
framework addresses.

The platform serves five stakeholder types whose requirements
are structurally conflicting. Farmers need simple low-bandwidth
interfaces and protection from surveillance by actors who
control their market access or could mandate livestock culling
in response to disclosed disease outbreaks. Government~(RAB)
needs aggregate analytics for disease surveillance, requiring
population-level data that farmers have direct interests in
withholding. Veterinarians need content moderation and
validation tools. NGOs and market actors need impact metrics
without extractive access to individual participation
histories. Administrators need system monitoring across all
stakeholder types.

Government disease surveillance and farmer protection from
surveillance consequences are not preferences that can be
reconciled: they are structural opposites. Architectural
choices determine whose needs are prioritized when conflicts
arise.

\subsection{Architectural Decisions with Governance Rationale}

All five framework mappings were instantiated. Farmer-driven
agenda is addressed through open Q\&A with community voting;
innovations from below through multi-signal reputation
aggregating peer validations, expert endorsements, and
contribution consistency, with weights documented in-app;
cross-level coordination through stakeholder-aligned
microservices separating farmer-facing services~(Content,
Quiz, Best Practices) from institutional services~(Analytics,
Moderation, Admin); and transparency for trust through an
explicit reputation formula visible in-app with tooltips.
The two decisions warranting fuller account are RLS and the
reputation interface visible to users.

\smallskip\noindent\textbf{Row-level security as governance
guarantee.} PostgreSQL RLS enforces data isolation at the
query layer: policies automatically filter results based on
user role, so government dashboards receive only aggregate
statistics from the same database farmers post to.
Application-level filtering relies on developer discipline
and fails with a single missed \texttt{WHERE} clause; RLS
rejects unauthorized queries at the database layer regardless
of application implementation. This makes the safe spaces
guarantee architectural rather than conventional, transferring
the trust boundary from application-layer discipline to
database-layer policy configuration.

\begin{figure}[t]
  \centering
  \includegraphics[width=\columnwidth]{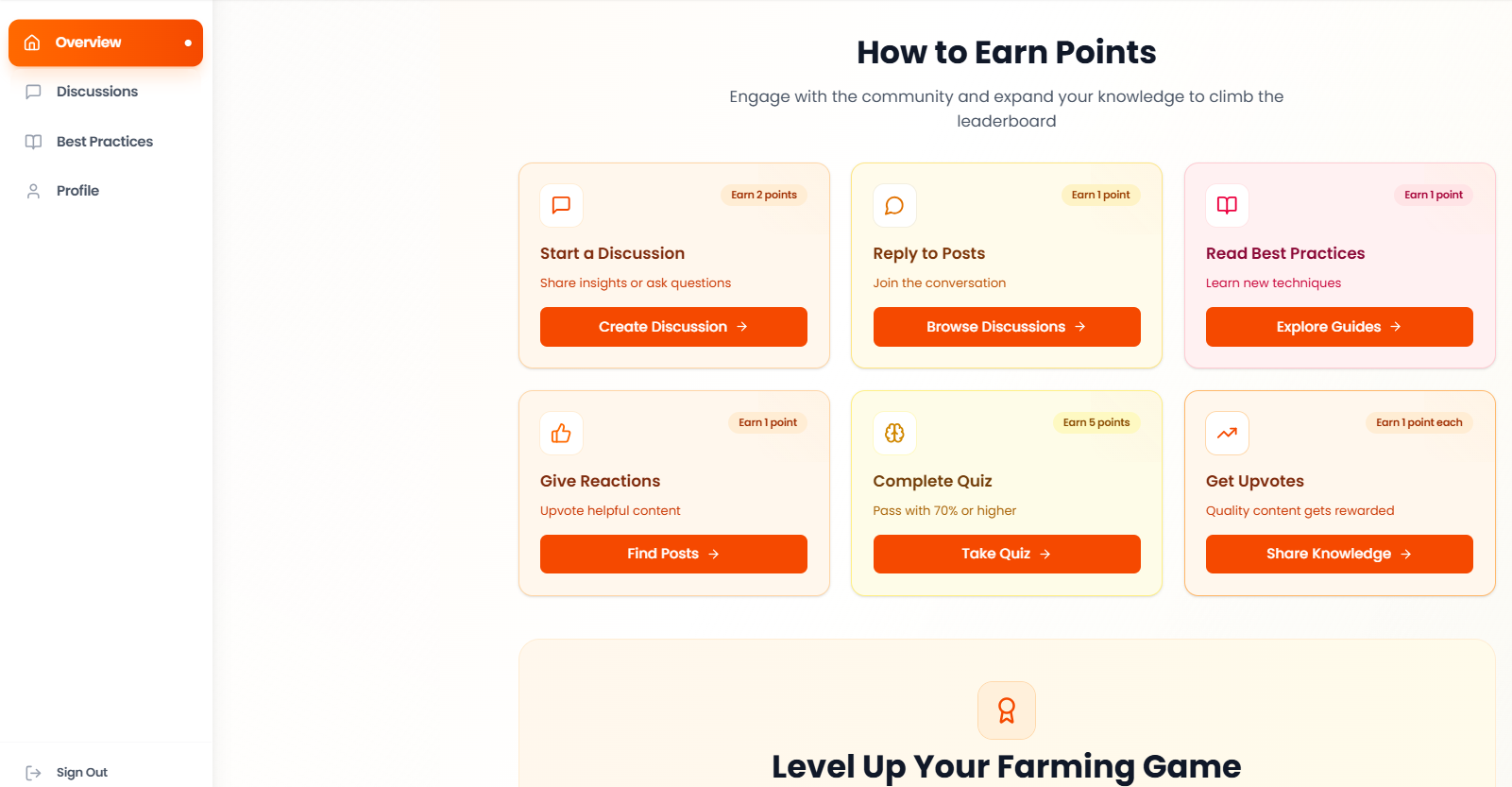}
  \caption{Gamification overview screen showing explicit point
    values for each action. Peer contribution actions (Start a
    Discussion, Give Reactions, Get Upvotes) are weighted
    alongside expert-defined activities, making the
    multi-signal reputation logic visible to users.}
  \label{fig:gamification}
\end{figure}

\smallskip\noindent\textbf{Reputation transparency in the
deployed interface.} Figure~\ref{fig:gamification} shows the
gamification overview screen with explicit point values for
each action: peer contributions (Start a Discussion, Give
Reactions, Get Upvotes) are weighted alongside
knowledge-consumption activities (Read Best Practices,
Complete Quiz), operationalising both transparency-for-trust
and innovations-from-below in the deployed interface.

\subsection{Architecture Overview}

\begin{figure*}[!t]
  \centering
  \begin{tikzpicture}[node distance=0.35cm and 0.25cm,
    svc/.style={
      rectangle, minimum width=2.0cm, minimum height=0.85cm,
      font=\scriptsize\sffamily, align=center,
      rounded corners=2pt, line width=0.5pt
    },
    gwbox/.style={svc, draw=gatewayborder, fill=gatewaycomp},
    fbox/.style={svc, draw=farmerborder, fill=farmercomp},
    ibox/.style={svc, draw=instborder, fill=instcomp},
    shbox/.style={svc, draw=sharedborder, fill=sharedcomp},
    ebox/.style={svc, draw=extborder, fill=extcomp},
    actor/.style={
      rectangle, draw=clientborder, fill=white,
      minimum width=1.4cm, minimum height=0.65cm,
      font=\scriptsize\sffamily, align=center,
      rounded corners=1.5pt, line width=0.4pt
    },
    dbcyl/.style={
      cylinder, draw=databorder, fill=datacomp,
      shape border rotate=90, aspect=0.2,
      minimum width=2.6cm, minimum height=0.9cm,
      font=\scriptsize\sffamily, align=center,
      line width=0.5pt
    },
    grp/.style 2 args={
      rectangle, draw=#1, fill=#2,
      rounded corners=3pt, line width=0.6pt,
      inner sep=6pt
    },
    grplbl/.style={font=\scriptsize\sffamily\bfseries, text=#1},
    sa/.style={-{Stealth[length=4pt,width=3pt]},
               line width=0.6pt, color=arrowgray},
    wa/.style={-{Stealth[length=4pt,width=3pt]},
               line width=0.6pt, color=arrowgray, dashed},
    ea/.style={-{Stealth[length=4pt,width=3pt]},
               line width=0.6pt, color=extborder, densely dotted},
    al/.style={font=\tiny\sffamily, text=lblgray, fill=white,
               inner sep=1pt, outer sep=0.5pt},
  ]

  \node[actor] (farmer) {Farmer};
  \node[actor, right=0.35cm of farmer] (vet) {Vet/Expert};
  \node[actor, right=0.35cm of vet] (govt) {Govt (RAB)};
  \node[actor, right=0.35cm of govt] (ngo) {NGO/Market};
  \node[actor, right=0.35cm of ngo] (admin) {Admin};

  \begin{scope}[on background layer]
    \node[grp={clientborder}{clientbg},
      fit=(farmer)(admin), inner sep=10pt,
      label={[grplbl=clientborder]above:Stakeholder Clients}
    ] (cf) {};
  \end{scope}

  \node[gwbox, below=1.1cm of govt] (apigw)
    {API Gateway\\{\tiny\itshape REST/GraphQL}};
  \node[gwbox, right=0.5cm of apigw] (wsgw)
    {WebSocket\\Gateway};

  \begin{scope}[on background layer]
    \node[grp={gatewayborder}{gatewaybg},
      fit=(apigw)(wsgw), inner sep=8pt,
      label={[grplbl=gatewayborder]above:Gateway Layer}
    ] (gf) {};
  \end{scope}

  \node[fbox, below=1.5cm of apigw, xshift=-0.3cm] (cs)
    {Content\\Service};
  \node[fbox, left=0.25cm of cs] (qs) {Quiz\\Service};
  \node[fbox, right=0.25cm of cs] (rs) {AI-Assist\\Service};
  \node[fbox, right=0.55cm of rs] (bps) {Best\\Practices};

  \begin{scope}[on background layer]
    \node[grp={farmerborder}{farmerbg},
      fit=(qs)(bps), inner sep=8pt,
      label={[grplbl=farmerborder]above:Farmer-Facing Services}
    ] (ff) {};
  \end{scope}

  \node[ibox, left=0.9cm of qs] (as) {Analytics\\Service};
  \node[ibox, left=0.3cm of as] (ms) {Moderation\\Service};

  \begin{scope}[on background layer]
    \node[grp={instborder}{instbg},
      fit=(ms)(as), inner sep=8pt,
      label={[grplbl=instborder]above:Institutional Services}
    ] (iif) {};
  \end{scope}

  \node[shbox, below=1.7cm of ms, xshift=0.5cm] (rbac)
    {User \&\\RBAC};
  \node[shbox, right=0.25cm of rbac] (notif) {Notification};
  \node[shbox, right=0.25cm of notif] (gam)
    {Gamification/\\Reputation};
  \node[shbox, right=0.25cm of gam] (trans)
    {Translation\\Adapter};
  \node[shbox, right=0.25cm of trans] (auth) {Auth\\Service};

  \begin{scope}[on background layer]
    \node[grp={sharedborder}{sharedbg},
      fit=(rbac)(auth), inner sep=8pt,
      label={[grplbl=sharedborder]above:Shared Services}
    ] (sf) {};
  \end{scope}

  \node[dbcyl, below=1.5cm of notif, xshift=-0.5cm] (db)
    {PostgreSQL\\+ RLS};

  \begin{scope}[on background layer]
    \node[grp={databorder}{databg},
      fit=(db), inner sep=10pt,
      label={[grplbl=databorder]above:Data Layer}
    ] (df) {};
  \end{scope}

  \node[ebox, below=1.5cm of trans, xshift=0.5cm] (cld)
    {Cloudinary};
  \node[ebox, right=0.25cm of cld] (gtr) {Google\\Translate};
  \node[ebox, right=0.25cm of gtr] (gem) {Gemini\\API};

  \begin{scope}[on background layer]
    \node[grp={extborder}{extbg},
      fit=(cld)(gem), inner sep=8pt,
      label={[grplbl=extborder]above:External APIs}
    ] (ef) {};
  \end{scope}

  \draw[sa] (farmer.south) -- ++(0,-0.5)
    -| ([xshift=-0.35cm]apigw.north west);
  \draw[sa] (vet.south) -- ++(0,-0.35)
    -| ([xshift=-0.2cm]apigw.north);
  \draw[sa] (govt) -- (apigw);
  \draw[sa] (ngo.south) -- ++(0,-0.35)
    -| ([xshift=0.2cm]apigw.north);
  \draw[sa] (admin.south) -- ++(0,-0.5)
    -| ([xshift=0.35cm]apigw.north east);
  \draw[wa] ([xshift=0.15cm]farmer.south) -- ++(0,-0.3)
    -| ([xshift=-0.15cm]wsgw.north);
  \draw[wa] ([xshift=0.15cm]vet.south) -- ++(0,-0.2)
    -| ([xshift=-0.4cm]wsgw.north);

  \draw[sa] (apigw.south) --
    node[al, right, xshift=1pt] {routes} (cs.north);
  \draw[sa] ([xshift=-0.3cm]apigw.south) -- (as.north);
  \draw[sa] ([xshift=0.4cm]apigw.south) -- ++(0,-0.3)
    -| (auth.north);

  \draw[sa] (cs.south) --
    node[al, right, xshift=2pt] {\tiny reputation} (gam.north);
  \draw[sa] ([xshift=-0.4cm]cs.south) -- ++(0,-0.3)
    -| node[al, pos=0.25, left] {\tiny alert} (notif.north);
  \draw[sa] ([xshift=0.3cm]cs.south) --
    node[al, right, xshift=2pt] {\tiny translate}
    ([xshift=-0.2cm]trans.north);
  \draw[sa] ([xshift=-0.3cm]ms.south) -- ++(0,-1.0)
    -| node[al, pos=0.12, left] {\tiny mod. alert}
    ([xshift=-0.5cm]notif.north west);
  \draw[sa] (rs.east) --
    node[al, above, yshift=1pt] {\tiny fetch practices}
    (bps.west);
  \draw[sa] ([xshift=0.3cm]as.south) --
    node[al, right, xshift=2pt] {\tiny verify role}
    ([xshift=0.2cm]rbac.north);

  \draw[sa] ([xshift=-0.15cm]cs.south) -- ++(0,-0.4)
    -| ([xshift=0.2cm]db.north);
  \draw[sa] (gam.south) -- ++(0,-0.15)
    -| ([xshift=0.5cm]db.north);
  \draw[sa] ([xshift=0.15cm]as.south) -- ++(0,-0.4)
    -| ([xshift=-0.2cm]db.north);
  \draw[sa] ([xshift=0.15cm]ms.south) -- ++(0,-0.55)
    -| ([xshift=-0.5cm]db.north);
  \draw[sa] (rbac.south) -- (db.north);

  \draw[ea] ([xshift=0.15cm]cs.south) -- ++(0,-0.25)
    -| (cld.north);
  \draw[ea] (trans.south) -- (gtr.north);
  \draw[ea] ([xshift=0.15cm]rs.south) -- ++(0,-0.25)
    -| (gem.north);

  \node[
    rectangle, draw=rlsborder, fill=rlsbg,
    rounded corners=2pt, line width=0.4pt,
    font=\tiny\sffamily, align=left,
    text width=4.2cm, inner sep=4pt,
    anchor=north west
  ] (rlsnote) at
    ([xshift=-0.5cm, yshift=-0.3cm]df.south west) {
    \textbf{RLS Policy Enforcement}\\[1pt]
    Farmer/Expert \quad$\rightarrow$ Individual posts\\
    Government \quad\;\;\,$\rightarrow$ Aggregates only\\
    NGO/Market \quad\,$\rightarrow$ Metrics, no PII\\
    Admin \quad\quad\quad\quad\;\,$\rightarrow$
      System + audit logs\\[2pt]
    \textit{Same queries, different results per role}
  };
  \draw[thin, color=rlsborder] (db.south)
    -- ++(0,-0.2) -| (rlsnote.north);

  \node[
    rectangle, draw=clientborder, fill=white,
    rounded corners=2pt, line width=0.4pt,
    font=\tiny\sffamily, align=left, text width=3.5cm,
    inner sep=4pt, anchor=north east
  ] (legend) at
    ([xshift=0.3cm, yshift=-0.3cm]ef.south east) {
    \textbf{Legend}\\[2pt]
    \rule[0.3ex]{0.7cm}{0.5pt}\;$\blacktriangleright$\;
      Internal sync.\ call\\[1pt]
    {\color{arrowgray} - - - -}\;$\blacktriangleright$\;
      WebSocket\\[1pt]
    {\color{extborder}$\cdots\cdots$}\;$\blacktriangleright$\;
      External API call\\[3pt]
    {\color{farmercomp}\rule{0.35cm}{0.2cm}}\;
      Farmer-facing\\[1pt]
    {\color{instcomp}\rule{0.35cm}{0.2cm}}\;
      Institutional\\[1pt]
    {\color{sharedcomp}\rule{0.35cm}{0.2cm}}\; Shared
  };

  \end{tikzpicture}
  \caption{System architecture showing stakeholder-aligned
    service decomposition and data isolation through
    row-level security}
  \label{fig:architecture}
\end{figure*}
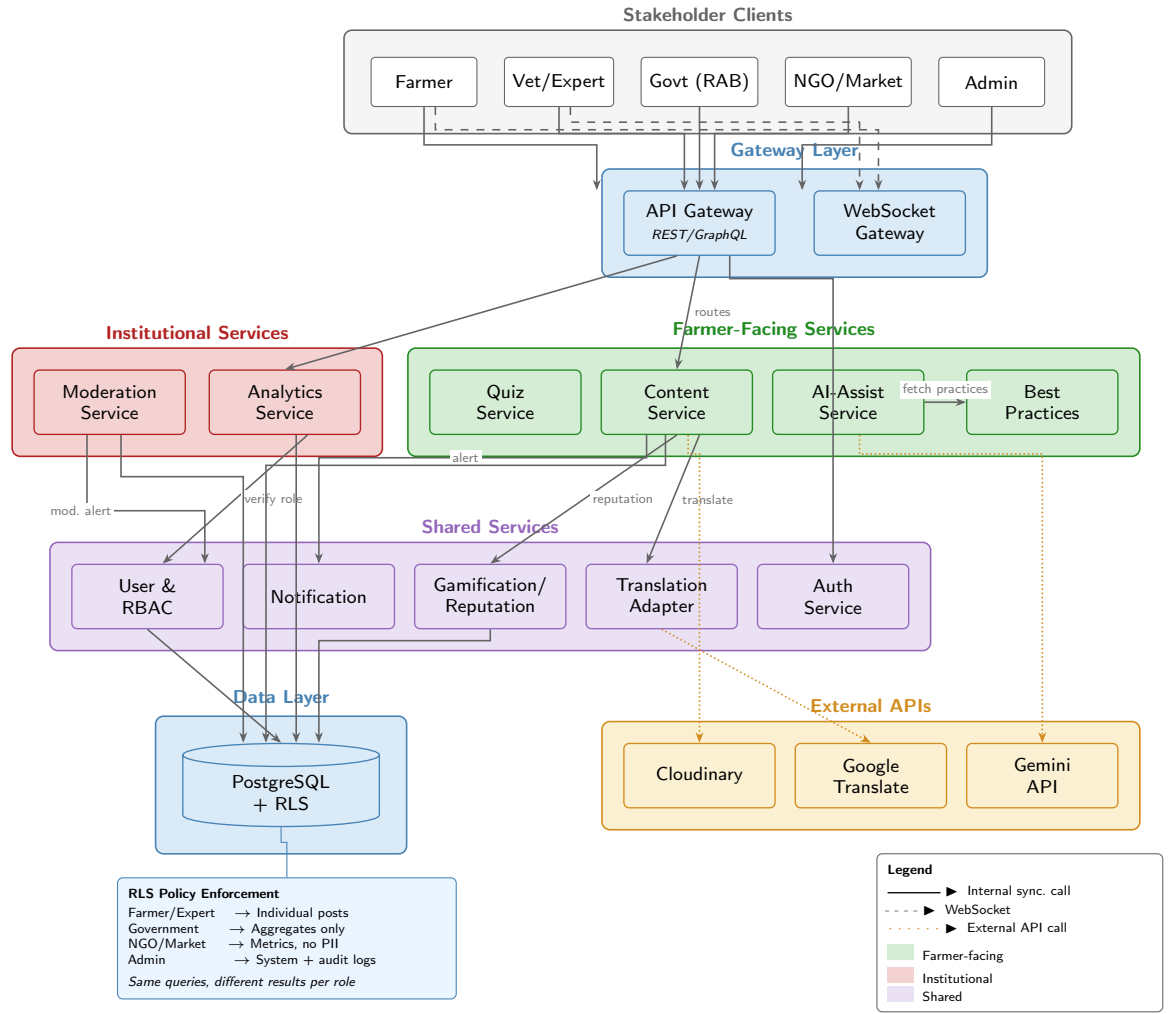

Figure~\ref{fig:architecture} maps each framework principle
to its architectural locus. Safe spaces is enforced at the
Data Layer through PostgreSQL RLS: the policy-enforcement
annotation shows how identical queries return role-specific
results (individual posts for farmers and experts, aggregates
for government, metrics without personally identifiable
information (PII) for NGOs and market actors, system and
audit logs for administrators). Cross-level coordination is
realized through stakeholder-aligned service decomposition,
visible as the Farmer-Facing Services group (Content, Quiz,
AI-Assist, Best Practices) separated from the Institutional
Services group (Analytics, Moderation), each independently
optimizable. Farmer-driven agenda resides in the Content
Service, which hosts the open Q\&A and community-voting
features described above. Innovations from below resides in
the Gamification/Reputation service among Shared Services,
where multi-signal aggregation lets peer recognition
accumulate without expert gatekeeping. Transparency for trust
is most directly visible in the user interface of
Figure~\ref{fig:gamification}; its architectural locus is the
same Gamification/Reputation service whose scoring logic
Figure~\ref{fig:gamification} surfaces.

\section{Discussion}
\label{sec:discussion}

Several limitations bound the framework's claims. The
framework was derived from one platform's architectural
decisions, introducing post-hoc rationalization risk:
governance-driven choices cannot be cleanly separated from
those mapped retrospectively to governance principles.
Single-context instantiation in pig farming in Rwanda means
the mappings in Table~\ref{tab:mapping} require validation
across other stakeholder configurations and regulatory
arrangements. Architecture satisfying governance requirements
is necessary but not sufficient for good governance outcomes;
the platform is constructed but governance outcomes remain to
be observed.

The four falsifiable predictions structuring the planned empirical validation state that: 1. safe-spaces architecture supports higher farmer
disclosure than shared-database alternatives; 2. multi-signal
reputation surfaces more local innovations than expert-only
validation; 3. stakeholder-aligned decomposition yields higher
farmer satisfaction than monolithic alternatives; and
4. transparent algorithms support higher user trust than opaque
ranking. These will be evaluated through a pre/post judgment
study administered to platform users across all five
stakeholder types prior to and following deployment.

\section{Conclusion}
\label{sec:conclusion}

Digital platforms increasingly mediate multi-stakeholder
coordination, but governance principles developed for
facilitated meetings do not automatically transfer to software
architecture. Engineers building these platforms make
governance decisions whether they recognize them as such or
not. We have proposed a governance-architecture correspondence
framework, illustrated in a constructed platform for pig
farmers in Rwanda, that makes these decisions explicit and
debatable. As work in progress, the framework awaits
validation through the pre/post judgment study outlined as
contribution (3). We invite researchers and practitioners to
apply the framework to other multi-stakeholder platforms and
report whether the mappings transfer, require modification,
or do not transfer.

\subsubsection*{Data Availability}

Architectural decision records, row-level security policy
specifications, and the reputation scoring algorithm
specification will be deposited in the ECSA Zenodo community
before camera-ready submission. The platform codebase is
currently being prepared for public release and will be made
available following completion of deployment preparation.

\subsubsection*{Acknowledgements}

Omitted for blind review.

\subsubsection*{Disclosure of Interests}

The authors have no competing interests to declare that are
relevant to the content of this article.


\end{document}